\documentclass[journal = jpclcd]{achemso}
\usepackage{amsmath}
\usepackage{soul}
\usepackage{xcolor}
\usepackage{graphicx}
\usepackage{caption,subcaption}
\usepackage{cleveref}
\usepackage{subcaption}
\usepackage[utf8]{inputenc}
 \usepackage{float}
 \usepackage{mathtools}
  \usepackage{mathbbol}
\usepackage{listings}

\usepackage{mymacros_11Sep20}

\newcommand{\absv}[1]{\bigl\lvert #1 \bigr\rvert}
\newcommand{\cl}{Cl$^-$}
\newcommand{\na}{Na$^+$}
\newcommand{\ch}{e}
\newcommand{\argmin}{\mathop{\mathrm{argmin}}\limits}  
\newcommand{\eps}{\epsilon}
\newcommand{\loss}{\mathcal{L}}

\graphicspath{./imm}

\author{Nicodemo Di Pasquale}
\email{nicodemo.dipasquale@manchester.ac.uk}
\author{Joshua D. Elliott}
\affiliation{Department of Chemical Engineering and Analytical Science, University of Manchester, Manchester M13 9AL, United Kingdom}

\author{Panagiotis Hadjidoukas}
\affiliation{IBM Research, Z\"urich, Switzerland}

\author{Paola Carbone}
\affiliation{Department of Chemical Engineering and Analytical Science, University of Manchester, Manchester M13 9AL, United Kingdom}

\begin{tocentry}
\includegraphics[width=1.\textwidth]{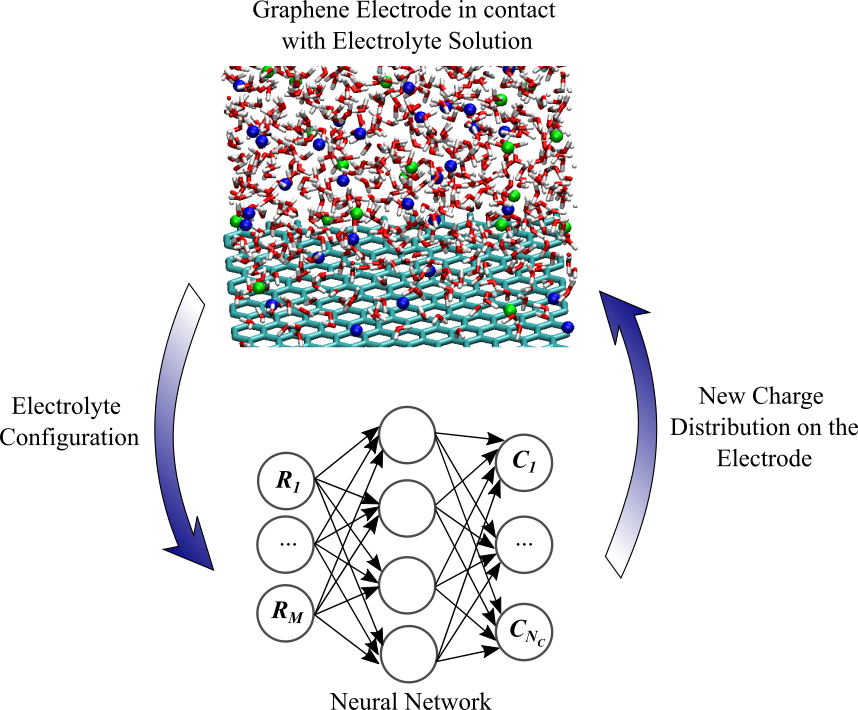}
\end{tocentry}

\title{Dynamically polarisable force-fields for surface simulations via multi-output classification Neural Networks}

\begin{document}

\begin{abstract}
    \noindent We present a general procedure to introduce electronic polarization into classical Molecular Dynamics (MD) force-fields using a Neural Network (NN) model. We apply this framework to the simulation of a solid-liquid interface where the polarization of the surface is essential to correctly capture the main features of the system. By introducing a multi-input, multi-output NN and treating the surface polarization as a discrete classification problem, for which NNs are known to excel, we are able to obtain very good accuracy in terms of quality of predictions. Through the definition of a custom loss function we are able to impose a physically motivated constraint within the NN itself making this model extremely versatile, especially in the modelling of different surface charge states.  The NN is validated considering the redistribution of electronic charge density within a graphene based electrode in contact with aqueous electrolyte solution, a system highly relevant to the development of next generation low-cost supercapacitors.
    We compare the performances of our NN/MD model against  Quantum Mechanics/Molecular dynamics simulations where we obtain a most satisfactorily agreement.
\end{abstract}

\maketitle


\noindent The first reports of Machine Learning (ML) in computational materials modelling emerged close to three decades ago \citep{Blank1995}, yet only very recently has their presence in this field become ubiquitous, in particular in the form of \textit{Supervised Learning} (SL)\citep{Goh2017,Noe2020,Zhang2020}. SL encompasses a group of methodologies with the same principal philosophy: given a set of observations in the form of input and output data, the goal is to determine a model that can make accurate output predictions given an arbitrary input. Examples of SL within materials modelling include Gaussian Approximated Potentials (GAP) \citep{Bartok2010,Bartok2015,Boussaidi2020} that use Gaussian random processes to predict atomistic Potential Energy Surfaces (PES) \citep{Behler2011,Behler2007,Schmitz2019}, Kriging regression \citep{DiPasquale2016JCTC, DiPasquale2016}, which is used in the geometrical optimization of molecules \citep{Zielinski2017} and PES prediction \citep{Davie2016JCC,Maxwell2016, DiPasquale2018}, kernel-ridge regression for the description of the multipole of a molecule \citep{Bereau2015JCTC,Scherer2020}, and Neural Networks (NN) that are also used to make predictions about the PES \citep{Behler2011,Behler2007,Schmitz2019} and predict the difference between forces obtained from DFT and classical force fields \citep{Pattnaik2020}. In particular, NNs represent the one most widely utilized techniques in materials modelling owing to their versatile and broad applications; NNs have been applied to the parametrization of wave functions \citep{Carleo2017} and quantum density matrices \citep{Hartmann2019} as well as applications within Quantum Monte Carlo simulations \citep{Fournier2020}. Here NNs are employed to model the quantum mechanical fluctuations of the electron density of a charged solid surface that lead to polarization effects at solid-liquid interfaces.

In the investigation of solid-liquid interfaces by classical Molecular Dynamics (MD) simulations, the standard practice in all-atom approaches is to assign to each species a fixed charge that is representative of its nuclear charge plus an attributable proportion of the average shared electron density. Polarisation effects that give rise to stronger attractive or repulsive interactions between non-bonded species may be treated in an average way through a parameterized non-bonded Lennard-Jones interaction potential \citep{williams_effective_2017, dockal_molecular_2019}. However, this treatment neglects the dynamical aspect of the interface polarizability which can be important for capturing the correct physisorprion and diffusion behaviour \citep{misra_uncovering_2021,misra_ion_2021}. In strictly metallic systems, the redistribution of the electronic density, and thence surface charge, can be modelled for instance by the constant potential method \citep{merlet_simulating_2013, wang_evaluation_2014, scalfi_charge_2020}. In semiconducting and insulating materials polarizable force fields can be applied to surfaces, accounting for dynamical effects by tethering a dummy charge to polarizable atoms via a harmonic spring, thereby allowing for modulation of the atomic charge density in response to the environment\cite{misra_insights_2017,pykal_ion_2019, misra_ion_2021}. In the case of graphene/electrolyte interfaces, in our previous work we observed that ions induce a long ranged redistribution of surface electron densities that can only be accurately accounted for by methods which compute directly the electronic surface density \citep{Elliott2020}. To this end we implemented an iterative Quantum Mechanics/Molecular Dynamics (QM/MD) workflow, by which the dynamics of surface-electrolyte interfaces can be modelled in a classical framework all the while including a QM description of the polarization of the surface. 

In the QM/MD scheme the state of the surface polarization evolves in response to the local electrostatic potential arising from the relative positions of molecules in the liquid phase. This could be for instance the water molecule dipole and/or charges associated with a solute. At a given time-step, the specific configuration of surface charges are obtained through Mulliken population analysis of the electronic charge density obtained at the Density Functional Tight Binding (DFTB) level of theory. In order to avoid very large-scale quantum mechanical calculations, only the surface atoms are treated by DFTB, and the specific arrangement of the electrolyte atoms enters into the calculation as a field of point charges. The DFTB surface atom populations are translated to a set of atomic charges and included as parameters in the classical MD force field (FF). Iteration of this procedure for many time steps ensures that there is feedback between the classically determined positions of the electrolyte atoms and the QM derived surface charges. 

Within this framework, the need of the DFTB calculations represents the bottleneck for the simulation time, and a trade-off between the accuracy and practical viability of the simulation must be established for the feedback between the QM and classical model. High frequency updates of the surface charge would improve the sampling of the electronic potential energy surface and reduce the time lag between the QM and MD calculations, but this comes at the cost of an enormous slow down in the simulation time. Supervised Learning of the QM polarizability, in particular using NNs, represents a novel avenue by which we can avoid the computationally expensive QM calculations, instead calling upon a trained model \emph{in-situ} to retain the dynamical description of the polarizability of the surface.  Along these lines, this work, introduces a NN model for the on-the-fly prediction of the surface atomic partial charges, within a classical force field (FF), that accounts for the evolving polarizability of the surface. The NN, which is trained over the QM calculations is then fully integrated into the ML/MD workflow replacing the QM calculations, still effectively reaching the same goal of obtaining an improved FF which is not constrained by fixed point charges. 

It is worth highlighting, our approach includes substantial deviations from the standard application of NN models to MD, more specifically: (\textbf{i})  We propose a \text{multi-output} neural network scheme \citep{Xu2019}, where a single NN gives for each prediction the instantaneous value of the charge on \textit{each} atom of the surface, (\textbf{ii}) We introduce a formal constraint in the generation of the NN model, to link the model to the physics of the system where the total surface charge must be preserved. This, in turn, is done by  modifying  the Loss Function (LF) used to train the model. %
Finally, (\textbf{iii}) we simplify the problem, from a standard regression one, where the prediction involves real numbers, to a problem similar to classifications where we are predicting (integer) classes instead. 
Although classification problems find use outside of computational materials science, they received little attention within the community. Here we show that they increase the flexibility and performances of the ML models particularly for the problem at hand (i.e. simulating a solid/liquid interface).  We validate the framework simulating the interfacial properties of an electrified graphene/electrolyte interface.

The system considered here is a charged semi-infinite graphene electrode in contact with a 1M NaCl electrolyte solution. The electrode is comprised of $N_C=336$ carbon atoms and carries an excess charge of 4 $e$. The electrolyte solution has 2065 water molecules and 90 and 86 fully dissociated Na$^+$ and Cl$^-$ ions respectively. A sketch of this system is presented in \cref{Fig:NN}. It should be noted that the excess of Na ions balances the charge of the electrode, preventing problems with the computation of long-ranged electrostatic interactions in the MD step. For a detailed description of the system and the model we refer to the Supporting Information (SI) (see Sec. S1 of the SI) and \citep{Elliott2020}. In our previous work, (see \citep{Elliott2020}) we observed that there exist a finite amount of charge for which any two charges differing by this value are not seen as different by the system. This observation, which will be made more quantitative in the next section, will be essential for the work developed here.

\section{Network Structure}

\noindent The first important novelty added in this work is that we can simplify  the problem by not considering a pure regression.  If the difference between two carbon partial charges is below a certain threshold, quantified as $\eps=0.015\, \ch $ \citep{Elliott2020}, the classical system is not able to distinguish between the different charges.  By looking at the distribution of the charges, dividing them into bins of finite size and assigning a label to each of them, we can ask our NN to predict the bin in which the particular charge falls. That means that we translated a pure regression problem into a labelling (classification) problem, even though a standard regression seems a natural choice given the continuity of the value of the charge.  The problem becomes to find the correct label for each surface carbon atom (i.e. correct class in the distribution of charges) given a specific electrolyte configuration, which in this case represents the input of our NN. Once the label is obtained, it can be mapped back to the corresponding charge. When a given charge is assigned to one of the bins it is replaced by the median value of that bin. Therefore, the size of the bin, $b$,  must be chosen smaller than $\epsilon$. However, we require a stricter condition on $b$, in particular we require that $b<\eps/2$. The reason lies in the discretization performed when we assign each charge to the bins, which is explained in detail in the SI (see Sec. S.2.1). If $b<\eps/2$ then a label prediction  which is close enough to the real one can be still considered correct, as will be shown in the next section. In this work we use a value of $b=0.007\,\ch$.

\begin{figure}[H]
\centering
\includegraphics[width=0.9\textwidth]{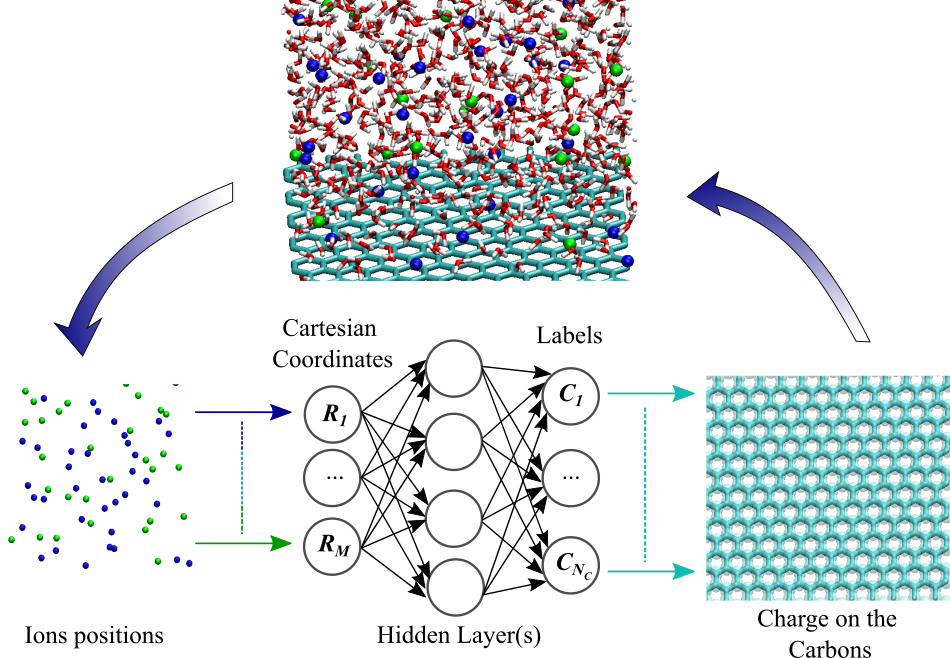}
\caption{Sketch of the loop for NN/MD calculations which shows the sequence of operations included. We highlighted the Multi-output Neural Network step with $M$ inputs and $N_C$ outputs, along with the definitions of the inputs extracted from the MD simulation (the position of the ions) and outputs entering into the MD simulation (i.e. the charge on the carbons). For more details see Sec. S.1 and Sec. S.3 of the SI.}
\label{Fig:NN}
\end{figure}

The basic architecture of a NN, ( see \cref{Fig:NN}) is represented by a set on neurons in several layers; each neuron within a layer is connected to all of the neurons in the subsequent layer. The first and last layers are special ones, the first layer represents the input layer, and the number on neurons here is equal to the number of features of the problem. The last layer is the output layer and in our case is composed of several outputs \citep{Borchani2015}. 

The output of the $j$-th neuron in the $k$-th layer, $y_j^{(k)}$, depends on the output of the previous layer $k-1$ and can be written as:
\begin{equation}
    y_j^{(k)} = f \cip{b^{(k-1)} + \sum_{i=1}^{m^{(k-1)} } w_{i,j}^{k} y_i^{(k-1)}}
\end{equation}
where the superscript ($k-1$) refers to objects in the $k-1$-th layer, $m^{\cip{k-1}}$ is the number of neurons in the layer, $w_{i,j}^{k}$ is the weight associated to the $i$-th neuron, $b^{(k-1)}$ is the bias and $f$ is the activation function.
If $y_j^{(k)}$ is the output of the last layer $k$ runs over all the different outputs. 
Here, we considered as activation function, $f$, the Rectified Linear Unit, ReLU which represents a good compromise between speed and robustness of the model.

In order to obtain a NN model, the weights $w_{i,j}^{k}$ must be optimized for each neuron in each layer, in practice this means minimising a certain \textit{Loss Function} (LF), $\loss(\bfy^{(out)}_i,\hat{\bfy}_i)$, which measures the ``distance'' between prediction and true value of the property:
\begin{equation}\label{Eq:argmin}
    \bfw^\mathrm{opt} = \argmin_\bfw \loss(\bfy^\mathrm{(out)},\hat{\bfy})
\end{equation}
where $\bfy^\mathrm{(out)}$ and $\hat{\bfy}$ are the $N_C$-dimensional vector of the respectively true and the predicted label attached to, in the present case, the $N_C$ carbons within the graphene layer. 
 

By definition, the true charges computed in the QM step sum to the charge applied to the electrode.  In order to enforce this constraint we include it as an extra term into the loss function appearing in \cref{Eq:argmin}. Our new loss function reads as:
\begin{align}\label{eq:customloss}
	\loss(\bfy^{(out)},\hat{\bfy}) & = \frac {1}{N_c} \left[  \sum_{i=1}^{N_C} \absv{ y^{(out),j}_i   - \hat{y}_i  } + \absv{ \sum_{i=1}^{N_C} \bfy^{(out)} \cdot \mathbb{1}   - \sum_{i=1}^{N_C} \hat{\bfy} \cdot  \mathbb{1} } \right ]
\end{align}
where $ \mathbb{1} $ is the $N_C$-dimensional vector of ones, and $(\cdot)$ is a scalar product. The first absolute value is the Mean Absolute Error (MAE) loss function optimized during the NN training, and represents the magnitude of error committed on the predictions. The second absolute value represents the penalty over the predictions calculated as the difference between the sum of the total predicted charges and that of the true total charge.


An important part of the creation of a ML model is the selection of the inputs, or \textit{features}. In our system, the distribution of charges on the graphene sheet depends on the configuration of the water molecules and ions in the electrolyte solution. However, the correlation among the positions of the water molecules and the ions during the simulation, strongly implies that we do not need to include all the molecules in the creation of the \textit{features}. In this work, we show that using the ions configurations is enough to obtain good descriptors for the training of the NN. This last fact represents a  key observation for the generation of NN models for MD simulations and here we argue that the amount of information needed to create good ML models can be reduced with respect to the naive choice of considering everything within the system. The number of feature we consider is $M=704$ calculated as: number of \cl~ + number of \na~ times four, i.e. the three spatial coordinates and the charge of each ion. 

Now, we need to explicitly describe the \textit{features} to be used in the NN model. Cartesian coordinates are not considered good candidates for ML models in MD. The fact that they lack some essential symmetries, i.e. they are not translational and rotational invariant as well as invariant to exchange of atoms, is generally a problem for ML model generation. The issue originates from the fact that NNs consider two input geometries which differ only by a rotation or translation of the system as being unique. In literature, different methods have been proposed to take into account these symmetries \citep{Davie2016,Ceriotti2020,Botu2015,Jinnouchi2020, Zuo2020}. 

In the present case, the use of absolute cartesian coordinates as input does not suffer of the problems mentioned above. In our work, the graphene carbons are fixed in space throughout the simulation. If two configurations  differ by a translation of any ion within the simulation box, they are different with respect to the fixed graphene interface. Therefore, they effectively represents different configurations \footnote{A translation symmetry along the direction perpendicular to the plane of the electrode exists but it's not considered here since all the configurations are obtained with respect the same position of the electrode.}.

We report all the details of the creation of the training set, the NN network and simulations in the Supporting Information (see Sec. S.2 of the SI). The NN are created by using the  \textit{Tensorflow/Keras} library v. 2.3.1 \citep{Abadi2016}.

The set up for the inclusion of the NN into the MD calculations is similar to the one reported in Sec. S.1 of the SI (see also \citep{Elliott2020} ) with the only difference being the replacement of the DFTB calculations with the prediction of the charges using the NN models. 

Every 5 ps the ions coordinate are extracted from the trajectory and transformed into the input configuration for the NN model, from which the new charges are predicted. A sketch of the loop of the NN/MD calculation is reported in \cref{Fig:NN}.

In practice, this functionality is implemented as a set of drivers which couples TensorFlow \citep{Abadi2016} with Gromacs \citep{Abraham2015}. Whilst external drivers slow the simulation time, they ensure that the developed models are fully transferable between different Classical molecular dynamics software packages.
 The fact that these scripts are not integrated directly within the code reduces the performances of the NN model which, however, remains well above the QM/MD ones in terms of computational time.


\section{Results}

\noindent In this section we will start by showing the performances of the NN model on a prediction set composed of 3000 electrolyte configurations and the resultant carbon charges computed by DFTB simulations (to which we will refer as the ``real'' charges). We then conclude by reporting the results of the fully integrated NN/MD simulation and comparing various properties  with those obtained from an analogous QM/MD trajectory.

\begin{figure}[htbp]
\begin{subfigure}{0.5\textwidth}
\begin{center}
\includegraphics[width=\textwidth]{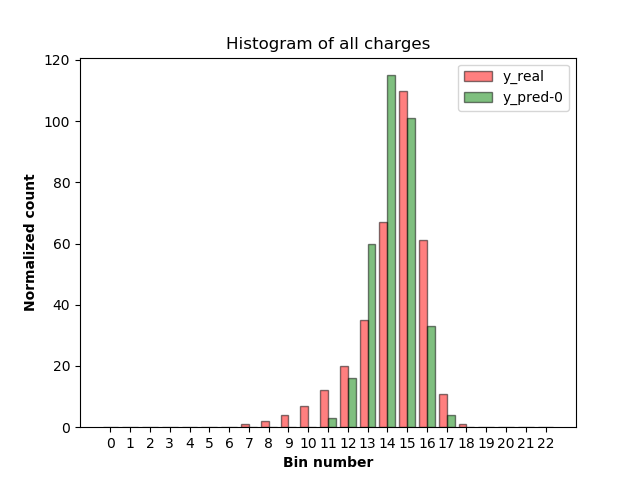}
\end{center}
\caption{\label{fig:H0}}
\end{subfigure}%
\begin{subfigure}{0.5\textwidth}
\begin{center}
\includegraphics[width=\textwidth]{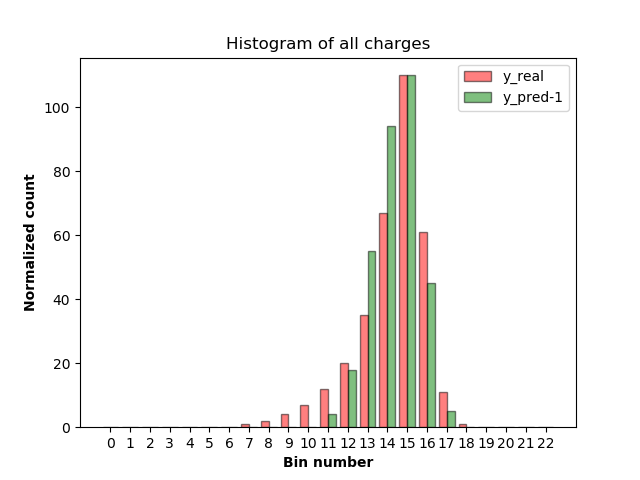}
\end{center}
\caption{\label{fig:H1}}
\end{subfigure}
\caption{Histogram of the true charges as calculated by the DFTB (in red) and comparison with their NN predicted values (in green). On the left panel we reported the charges as they are predicted from the NN, on the right panel we report the filtered charges using the threshold defined in the main text. }
\label{Fig:histo}
\end{figure}

In \Cref{Fig:histo}  we plot histograms that compare the distributions of the real and predicted charges.
Leveraging that  $i)$ the differences in the charges which are smaller than $\epsilon$ are not seen as different and $ii)$ the size of the bin, $b$, is such that $b<\epsilon/2$, it follows that if the NN predicts a label for a charge which is $\pm1$ away from its correct one, it can be assigned back to its correct label. By filtering out the results in \cref{fig:H0}, which represents the distribution as obtained by the NN model, using this consideration, we obtain the histogram shown in \cref{fig:H1}.
Our model is able to correctly capture the behaviour of the system in terms of identification of the most important classes, which in turn, represent the most likely observed charges. However, our model yields a lower likelihood that charges on the left tail of the distribution will be observed. These charges are those appearing with the least frequency during the time evolution of the electrolyte configurations, which are therefore likely to be under-represented using the random sampling we employed to construct the training set. An improved sampling of the training set may help in reducing this effect (e.g. \citep{Gastegger2017}), but as we will show next, under representation of these labels does not have a noticeable effect on the NN/ML simulation results.

\begin{figure}[htbp]
\centering
\includegraphics[width=9cm, height=6.5cm]{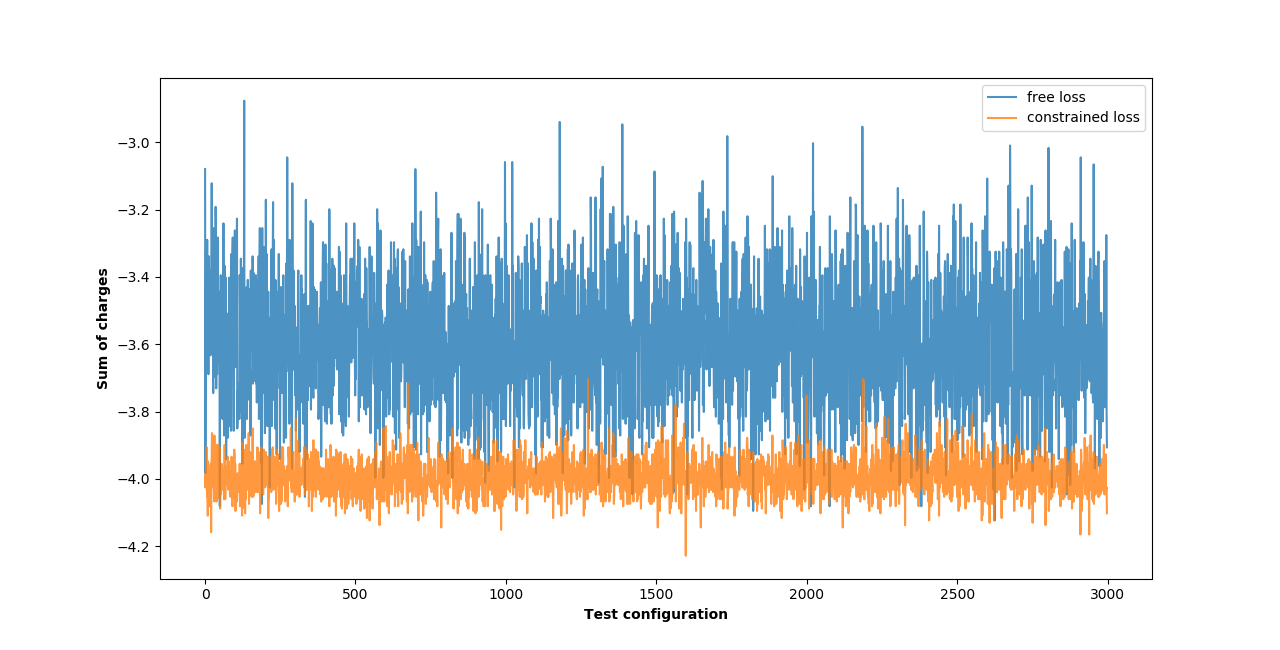}
\caption{Sum of charges for the test configurations}
\label{Fig:sumcharges}
\end{figure}

\Cref{Fig:sumcharges} reports the sum of the charges, for each frame, of the predicted set, with and without the constraint applied in the loss function. This serves to highlight the importance of the constraint since without it the sum of the predicted charges has a mean value different from the one set in the classical step (4.0 $e$ for the system considered here). If the NN does not conserve the electrode charge, then firstly, the modelled system is fundamentally different from the real system and secondly, on a more technical note this can lead to instabilities in the evaluation of the Ewald summation during the classical MD step, where the overall electrolyte charge no longer counter balances the electrode. In fact, our results suggest that the inclusion of a physically motivated terms within the loss function can lead to better models generally. 

We have shown in \cref{Fig:histo} and \cref{Fig:sumcharges} the performance of the predictions in terms of the relative frequency of the charges compared with the real ones.

On top of their histograms, we can also consider the real space distributions of the predicted labels in comparison with the true charges, since this will give rise to the dynamical feedback with the electrolyte during the NN/MD loop. In particular, this difference between the QM and NN charges should be minimal in order to avoid nonphysical charge (de)localization.

\begin{figure}
\centering
\includegraphics[width=0.5\columnwidth]{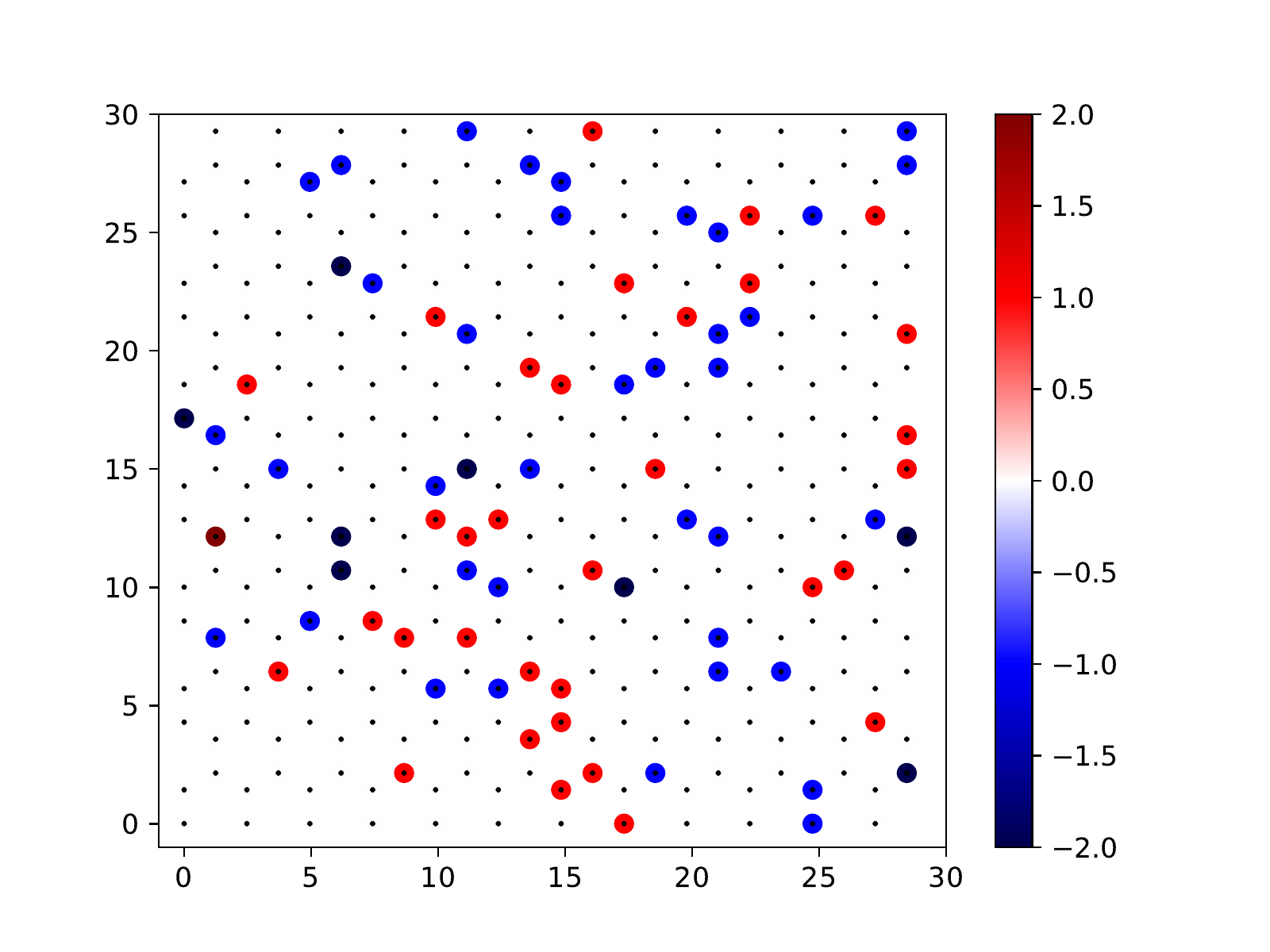}
\caption{A two dimensional plot of the difference between the QM and NN distribution of charges on the Carbon atoms in the graphene sheet for a randomly selected snapshot, a movie over the trajectory is available as part of the SI. The legend is given in units of the threshold $\epsilon$. }
\label{Fig:charges}
\end{figure}

 The comparison between the true and predicted charges has been carried out on the test set (see Sec. S.2 of the SI) by calculating the difference between the value of the real charges and the prediction of the NN model on the same configuration,  which we report in \cref{Fig:charges}. If the error on the charge is smaller than the threshold of $0.015$ than an error of zero is assigned to that particular carbon. We observe that given this constraint the difference between the real and predicted charges is zero for the majority of C atoms across all frames. Moreover, where individual C atoms take a value different from zero, the prediction appears to be in isolated in space and across different frames. As a consequence, the resultant polarization of the sheet is not affected and the contribution of these larger deviations is averaged out over the course of even several tens of ps.
As observed for \cref{Fig:histo}, the NN has a slight bias towards highest labels which can be possibly mitigated by a more accurate selection of the training set points, but overall the qualitative behavior is similar to the one observed in QM calculations with regions with a larger (negative) charge, regions mostly neutral and very few positive charges. 

\begin{figure}[htbp]
\centering
\includegraphics[width=9cm, height=6.5cm]{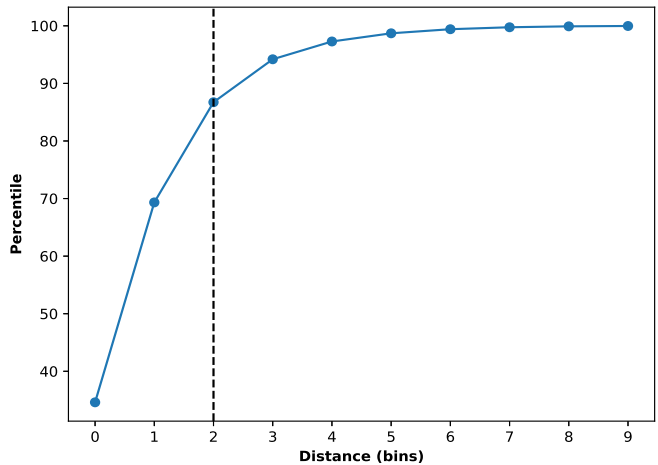}
\caption{S curves showing the distance of the predicted bin with respect the correct value. The vertical black dashed line represents the threshold $\epsilon$. }
\label{Fig:splot}
\end{figure}

\begin{figure}[htbp]
\centering
\includegraphics[width=9cm]{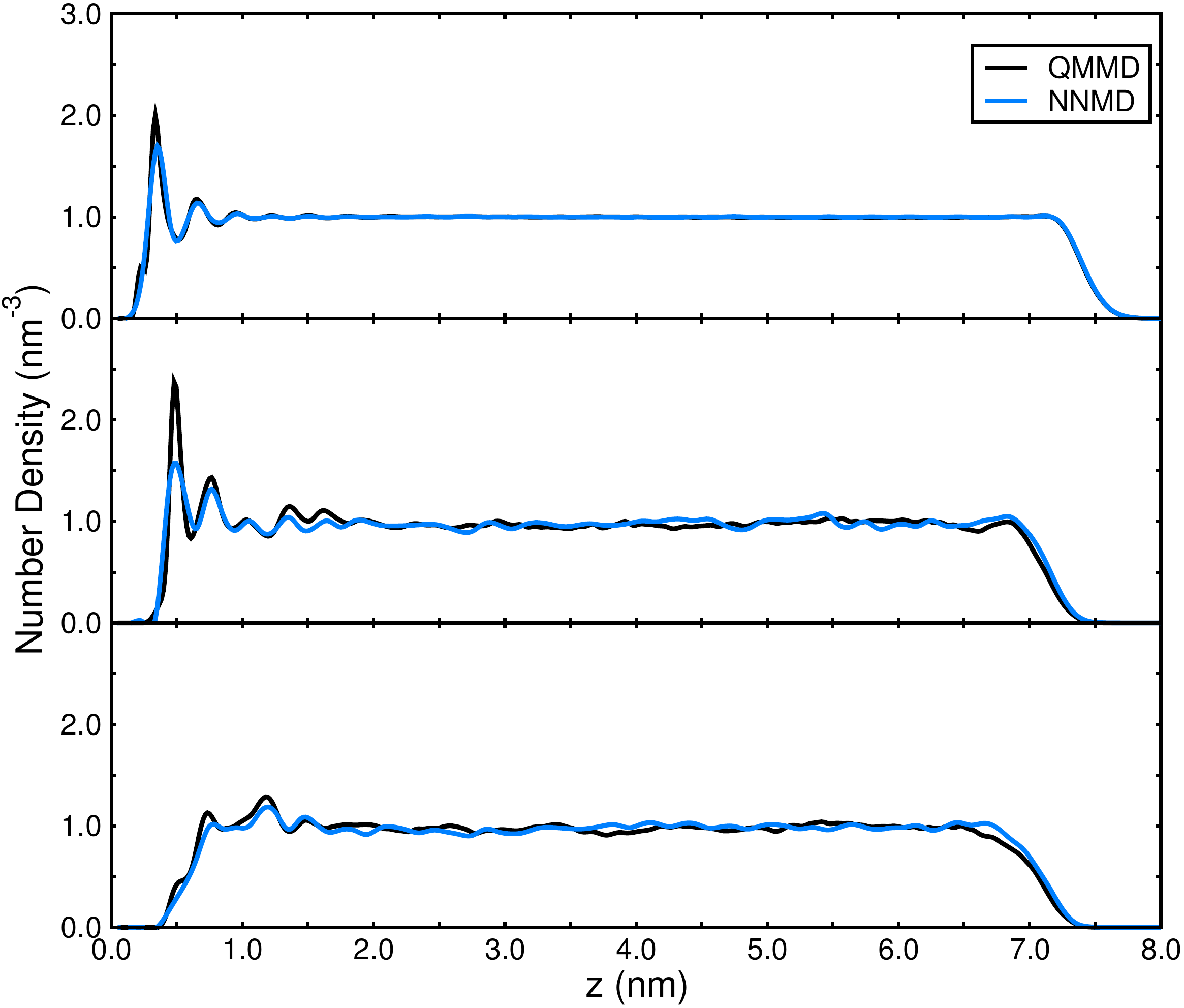}
\caption{From top to bottom: normalized density of water, \na~and \cl~as function of the distance from the graphene layer for QM/MD, NN/MD calculations.}
\label{Fig:density}
\end{figure}

The distribution of the predicted charged gives the overall behaviour of the predictions, but does not give any indication on the error committed in each prediction. The most straightforward evaluation of the performances of any NN is the prediction error with respect the charges on the prediction set. As  shown in  \cref{Fig:splot}, where we report an $S$-curve showing the percentile on the $y$-axis and the absolute error on the $x$-axis. Each point gives on the $y$-axis the percentage of the carbons in the predictions set with error lower than the one marked by the $x$ position. The error is given in terms of the distance of the predicted label from the true one. A distance of zero means the label was correctly predicted. The $S$-curve for the predictions on the charges on the graphene layer is plotted in \cref{Fig:splot}. Even though each prediction gives all the charges on the graphene layer at the same time, we consider for \cref{Fig:splot} each charges separately, i.e. the picture is showing the errors committed on a single charge. 
In this figure we also included a black vertical line at $0.015\,\ch$.  From \cref{Fig:splot} it results that with the $0.015\,\ch$ threshold we can consider correct almost 87\% of the charges predicted. Naturally, the threshold we used has still to be tested in a simulation, where we can confirm that such an approximation is enough to obtain reliable results from the MD simulations. 
\Cref{Fig:density} shows the normalized density of water, \na and \cl~across the simulation box as function of the distance from the graphene layer (which, in our configuration is perpendicular to the $z$-direction and is located at $z=0$). 
As expected, the density of the \cl~ion is smaller than the \na near the surface since the graphene layer is negatively charged. One thing we can notice from \cref{Fig:density} by comparing the relative height of the first peaks for water and sodium (comparable to the first solvation shell for the graphene layer) is that the relative height of the peaks is preserved in NN/MD. The distribution of the chlorine shows a better agreement with QM/MD calculations than the sodium.  This fact could be due to the fact that chlorine being, on average, repelled from the interface is less sensitive to the small differences between the QM and NN description of the graphene layer. 

These results show the potential of this different paradigm for NN for MD simulations. In particular, we showed how the classification problem can be used in the generation of NN for MD calculations instead of the more complicate regression one. We showed that Cartesian coordinates can be used as features, giving good results, even though other feature selection may improve the predictions as well as a more clever choice of the training set geometries.
In terms of speed-up of the calculations, the time needed for a NN/MD simulation is of the same order of magnitude of a standard MD simulation, which is a huge computational advantage compared with any other procedure to include surface polarizability. However, the model presented here suffers of a not fully integration with the MD code (see Sec. S.3 of the SI), which surely represents the next step in the development of NN/MD models.

\section{Acknowledgment}

The authors thank the European Union’s Horizon 2020 research and innovation programme project VIMMP under grant agreement no. 760907. The authors would like to acknowledge the assistance given by Research IT and the use of The HPC Pool funded by the Research Lifecycle Programme at The University of Manchester.

\bibliography{bibliography_20Feb20}

\end{document}